\documentclass[aps, prc, article, amsmath, amssymb, graphicx, longbibliography]{revtex4-1}

\usepackage{bm}
\usepackage{graphicx}
\usepackage{epsfig}
\usepackage{amsthm}
\usepackage{longtable}
\usepackage{booktabs}
\usepackage{epstopdf}
\usepackage[mathlines]{lineno}

\usepackage{grffile}
\usepackage{float}

\usepackage{amsthm}

\begin{document}

\title{A generalization of the Levinson theorem \\about the asymptotic value of the scattering phase shift%\footnote{to appear in Yad.~Fiz.~\textbf{83} (2020).}
}

\author{M. I. Krivoruchenko}
\address{National Research Center $^{\prime}$Kurchatov Institute$^{\prime}$ - ITEP \\ Cheremushkinskaya 25, 117218 Moscow, Russia}

\author{K. S. Tyrin}
\address{Kurchatov Institute, National Research Center \\ Kurchatov Square 1, 123182 Moscow, Russia}

\begin{abstract}
%In quantum scattering theory, the relationship between the difference in scattering phase shifts at threshold and infinity and the number of bound states is established by the Levinson theorem. 
In quantum scattering theory, there exists a relationship between the difference in the scattering phase shifts at threshold and infinity and 
the number of bound states, which is established by the Levinson theorem.
The presence of Castillejo, Dalitz and Dyson poles in the scattering amplitude, as well as Jaffe and Low primitives,
corresponding to zeros of $D$ function on the unitary cut, modify the Levinson theorem.
%The difference in scattering phase shifts at threshold and infinity is shown to be determined by the number of bound states, the number of Castillejo, Dalitz and Dyson poles, and the number of primitives.
The asymptotic value of the scattering phase shift is shown to be determined by the number of bound states, the number of Castillejo, Dalitz and Dyson poles, and the number of primitives.
Some consequences of the generalized theorem with respect to properties of nucleon-nucleon interactions are discussed.
\end{abstract}

\maketitle

Analyticity of the $S$ matrix is considered to be a consequence of micro-causality \cite{Chew:1968}. In the framework of perturbation theory, singularities of scattering amplitudes over kinematic invariants are determined by the Landau rules \cite{Landau:1959,Eden:1966}. For particles with momenta $p_1$ and $p_2$, the $S$ matrix in each partial wave is an analytic function of $s = (p_1 + p_2)^2$ on the physical sheet of the Riemann surface, except for simple poles corresponding to bound states, the left cut and the unitary cut. One of the consequences of the analyticity is the Levinson theorem \cite{Levinson:1949},
which relates the number of bound states with the scattering phase difference at threshold and infinity.
There is a generalization of this theorem to relativistic case \cite{Warnock:1963,Hartle:1966},
which takes into account multichannel structure of the $S$ matrix and CDD poles, introduced by Castillejo, Dalitz, and Dyson \cite{Castillejo:1956, Dyson:1957} to demonstrate the ambiguity in solutions of the Low equation \cite{Low:1955}.

The generalizations considered to date do not cover systems with the so-called "primitives", which appear in scattering theory as $P$-matrix poles. According to Jaffe and Low  \cite{Jaffe:1979}, multi-quark states correspond to poles of $P$ matrix, not $S$ matrix.
In the framework of the $N/D$ method, primitives manifest themselves
as zeros of $D$ function on the unitary cut, in which the scattering phase difference $\delta(s) - \delta(s_0)$ vanishes modulo $\pi$ with a negative slope.
CDD poles correspond to zeros modulo $\pi$ of the scattering phase difference with a positive slope.
Primitives allow to interpret nucleon-nucleon repulsive core in terms of the $s$-channel exchange of 6-quark states \cite{Simonov:1979,Simonov:1982,Simonov:1984,Bhasin:1985,Fasano:1987,Bakker:1994,Krivoruchenko:2010,Krivoruchenko:2011b,Krivoruchenko:2012ii}.
The $s$-channel exchange models are successful in describing nucleon-nucleon scattering \cite{Simonov:1979,Simonov:1982,Simonov:1984},
few-nucleon systems \cite{Bakker:1994},
pairing gap in neutron matter \cite{Krivoruchenko:2017} and
provide useful hints for searching narrow dibaryons \cite{Krivoruchenko:2011}.
A generalization of the Levinson theorem is of interest for systems in which primitives are identified.

Equation of state (EoS) of nuclear matter is important for astrophysics of compact objects.
The stiffest EoS is considered to be the one with the speed of sound $ a_s $ is equal to the speed of light $ c$.
In mean field models, this condition is satisfied asymptotically with increasing nuclear density,
given that nucleons interact through the $t$-channel exchange of the $\omega$-mesons \cite{Zeldovich:1961}.
The discovery of neutron stars with masses of about $2M_ {\odot}$ \cite{Demorest:2010, Antoniadis:2013}
excludes a wide class of soft EoS of nuclear matter based on the $t$-channel exchange models.
When repulsion dominates, the scattering phase decreases with increasing energy.
The stiffness of EoS is thereby sensitive to the asymptotic value of the scattering phase,
which, according to the Levinson theorem, is determined by the number of bound states and
depends, as we argue below, on the number of CDD poles and the number of Jaffe-Low primitives.

The one-channel $S$ matrix in a fixed partial wave can be expressed in terms of the scattering phase, $\delta(s)$,
or the Jost function $D(s)$:
\begin{equation} \label{S}
S = e^{2i\delta(x)} = \frac{D_{II} (s)}{D_{I} (s)}.
\end{equation}
$D_I (s)$ matches $D(s)$ on the physical (first) sheet of the Riemann surface to which the upper edge of the unitary cut $(s_0,+\infty)$ belongs.
Analytical continuation of $D_{I} (s) $ through the unitary cut
to the region $ \Im s < 0 $ leads to the non-physical (second) sheet of the Riemann surface.
The function $ D_{II} (s)$ is defined by analytical continuation of $D(s)$ from the lower edge of the unitary cut.
In the domain $ \Im s < 0$, when the variable $ s $ belongs to the physical sheet of the Riemann surface,
$D_{II} (s) $ coincides with $D(s)$.

$D(s)$ is analytic in the complex $s$ plane with the cut $(s_0,+\infty)$, the threshold $s_0$ is a branch point. The imaginary part of $D(s)$ on the unitary cut defines the $ N $ function. $D(s)$ has no zeros in the complex $s$ plane except for simple zeros on the real axis for $s < s_0$, which correspond to bound states, and simple zeros for $s > s_0$, which correspond to primitives. CDD (simple) poles are located on the real axis at $s \lessgtr s_0$. CDD poles permeate all sheets of the Riemann surface. Condition for the existence of a primitive, $D_{I}(s) = 0$ for $s \in (s_0,+\infty)$, is necessary and sufficient for $D_{II} (s) = 0$ because $D_{I}(s)$ and $D_{II} (s)$ differ only in the imaginary part, which in this case is equal to zero.

We consider the integral
\begin{equation}
J = \int^{+\infty}_{s_{0}}ds\ln(S)^{\prime} = 2i \int^{+\infty}_{s_{0}}ds\delta(s)^{\prime} = 2i(\delta(+\infty)-\delta(s_{0})).
\end{equation}
The integration is performed along the real axis. $J$ can be writen in terms of the Jost function:
\begin{equation}
J =\int^{+\infty}_{s_{0}}ds\ln\left( \frac{D_{II}(s)}{D_{I}(s)} \right)^{\prime} = \int^{+\infty}_{s_{0}}ds\left( \frac{D_{II}(s)^{\prime}}{D_{II}(s)}-\dfrac{D_{I}(s)^{\prime}}{D_{I}(s)} \right).
\end{equation}
Under the sign of the second integral, each of the summands has simple poles generated by CDD poles and primitives.
The poles and the zeros are cancelled in ${D_{II}(s)}/{D_{I}(s)}$, so there exists a region $ \Im s \lesseqqgtr 0$, adjacent to the unitary cut,
which is domain of analyticity of the integrand function.
The integration contour can be moved down, then the integral can be split into two terms:
\begin{equation} \label{4}
J = \int_{C}ds\dfrac{D_{II}(s)'}{D_{II}(s)} + \int_{C^{\prime}}ds\dfrac{D_{I}(s)'}{D_{I}(s)}.
\end{equation}
After the contour is shifted down, the argument of the function $D_{II} (s) $ belongs to the first ($I$) sheet of the Riemann surface.
The argument of the function $D_{I} (s)$ goes under the cut and arrives at the non-physical sheet $II$.
The integration paths are shown in Fig.~1. The contour $C$ lies on the physical sheet below the unitary cut,
while the contour $C^{\prime}$ lies under the unitary cut on the second sheet, where zeros of $D(s)$ corresponding to resonances are located.
We consider a fairly small contour shift so that $C^{\prime} $ did not intersect with zeros of $D(s)$ corresponding to resonances.

\begin{figure} [t]
	\begin{center}
		\includegraphics[angle = -90,width=0.618\textwidth]{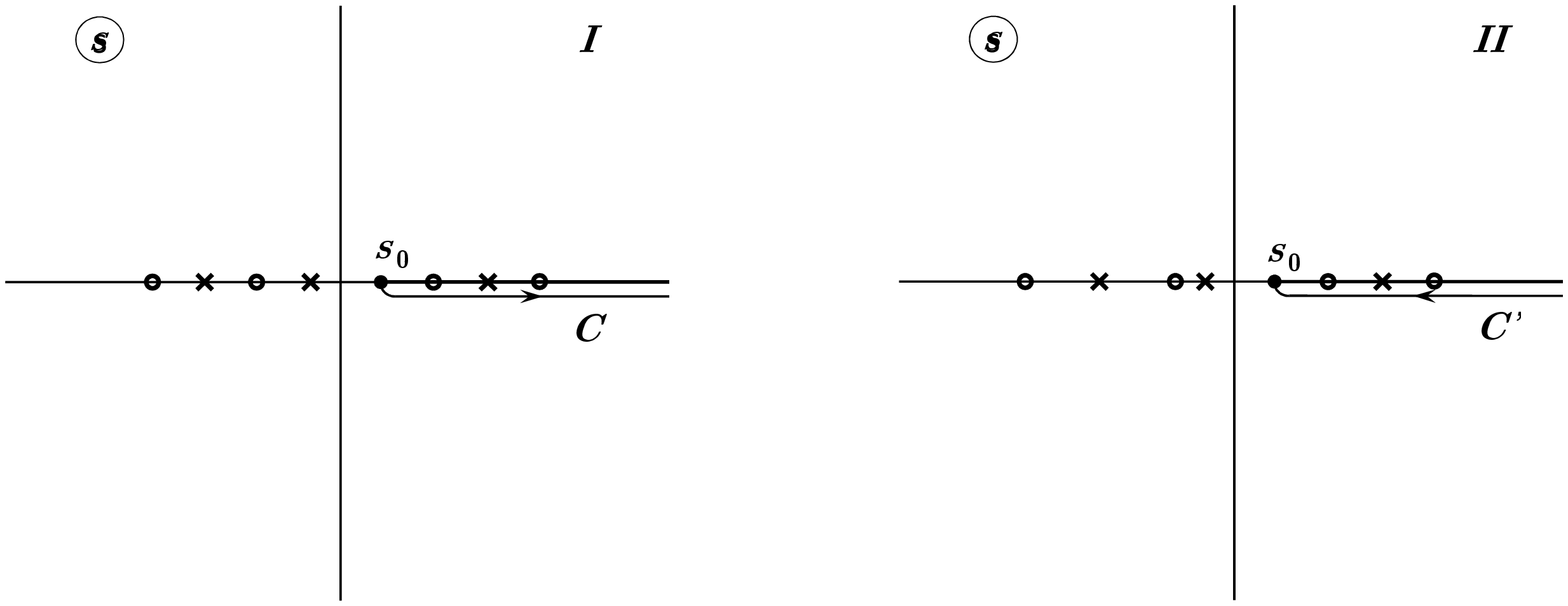}
\caption{Zeros and poles of the Jost function on the first ($I$) and second ($II$) sheets of the Riemann surface. Zeros (circles) correspond to bound states for $ s < s_0$ and primitives for $s > s_0$; poles (crosses) correspond to CDD poles. The arrows on the contours $ C $ and $ C^{\prime}$ indicate direction along which the integration is performed in Eq.~(\ref{4}).}
		\label{fig1}
	\end{center}
\end{figure}		

On the real axis, the integrand functions of Eq.~(\ref{4}) have simple poles corresponding to zeros of $D_{I} (s)$ and $D_{II} (s)$:
zeros of $D_{I} (s)$ for $s < s_0$ describe bound states.
These zeros usually do not match zeros of $D_{II} (s)$ for $s < s_0$ on the second sheet.
Zeros on the unitary cut $ (s_{0},+\infty) $ correspond to primitives. CDD poles belong to the real axis.
Zeros corresponding to primitives and CDD poles coincide on the Riemann sheets $I$ and $II$.
The integrand functions of Eq.~(\ref{4}) have $s_{0}$ as the branch point.
It is assumed that in a neighborhood of $s_{0}$ $D (s)$ is bounded,
so that the integral around $s_0$ vanishes when the radius of the circle tends to vanish.

We deform the contour $C^{\prime}$ through the unitary cut and end up on the sheet $I$, as shown in Fig.~2. During the deformation process,
there appears a contribution to the contour integral from the residues:
\begin{equation} \label{1}
J_1 = -2\pi i\sum_{i=1}^{n_{\mathrm{p}}}\mathrm{Res} (\frac{D_{I}(s)'}{D_{I}(s)},s_i) - 2\pi i\sum_{j=k+1}^{n_{\mathrm{CDD}}}\mathrm{Res} (\frac{D_{I}(s)'}{D_{I}(s)},s_j),
\end{equation}
where $ n_{\mathrm{p}}$ is the number of primitives, $ k $ is the number of CDD poles below the threshold and $ n_{\mathrm{CDD}} $ is the total number of CDD poles.
On the first sheet, we perform the integration along the contours $ C $ and $ C^{\prime}$ and
add the integral along an infinitely distant circle, $ C_{\infty} $,
as shown on the left panel of Fig.~2.
A sufficient condition for the integral over $ C_{\infty} $ to vanish is the condition $s D(s)^{\prime}/D(s) \to 0$ for $|s| \to \infty$.
This condition is assumed to be met.

The contour $\Gamma = C \cup C^{\prime} \cup C_{\infty}$ is closed.
Around the unitary cut for $ \Im s \lesseqqgtr 0$ $D_{I} (s)$ is analytical extension of $D(s)$, so that in Eq.~(\ref{1}) $D_{I} (s)$ can be replaced by $D(s)$.
$D_{I} (s)$ on the upper edge of the unitary cut and for $\Im s > 0$ and
$D_{II} (s)$ on the lower edge of the unitary cut and for $\Im s < 0$
%of the physical sheet $I$
are analytical extensions of $D(s)$.
$D_{I} (s)$ and $D_{II} (s)$ entering the contour integral over $ \Gamma $ can therefore be replaced with $D(s)$ to give
\begin{eqnarray} \label{2}
J_2=\oint_{\Gamma}\frac{D(s)'}{D(s)}ds=-2\pi i \sum_{l=1}^{n_{\mathrm{b}}}\mathrm{Res}(\frac{D(s)'}{D(s)},s_l) - 2\pi i \sum_{j=1}^{k}\mathrm{Res}(\frac{D(s)'}{D(s)},s_j),
\end{eqnarray}
where $ n_{\mathrm{B}}$ is the number of bound states.
In a neighborhood of bound state $D(s) \sim (s-s_{l})$,
in a neighborhood of primitive $D(s) \sim (s-s_ {i})$, while
in a neighborhood of CDD pole $D(s) \sim 1/(s-s_{j})$,
as a result
$\mathrm{Res}({D(s)^{\prime}}/{D(s)}, s_l) = \mathrm{Res}({D(s)^{\prime}}/{D (s)},s_i) = - \mathrm{Res} ({D(s)^{\prime}}/{D (s)},s_j) = 1$.
Given that $ J = J_1 + J_2$, we find
\begin{equation}
J=-2\pi i (n_{\mathrm{b}} + n_{\mathrm{p}} - n_{\mathrm{CDD}}),
\end{equation}
and finally,
\begin{equation} \label{Levinson}
\delta(+\infty)-\delta(s_{0})=-\pi (n_{\mathrm{b}} + n_{\mathrm{p}} - n_{\mathrm{CDD}}).
\end{equation}

\begin{figure} [t]
	\begin{center}
		\includegraphics[angle = -90,width=0.618\textwidth]{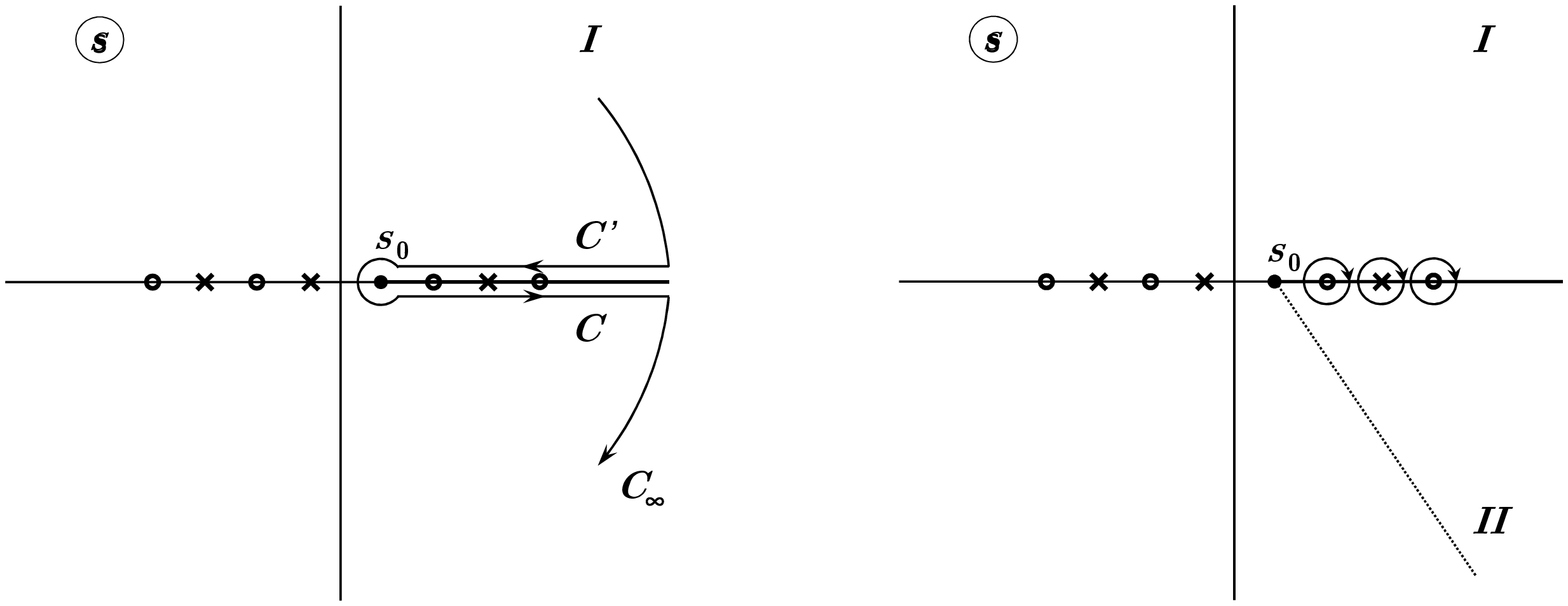}
		\caption{
The left panel shows the physical sheet $I$ of the Riemann surface of the function $D(s)$ with
the contours $C^{\prime}$ and $C$ connected by an arc bypassing threshold $s_0$.
At positive infinity, both contours are continued along the circle $C_ {\infty}$ with a radius tending to infinity.
The right panel shows a part of the fourth quadrant of the physical sheet, bent along the dotted line starting from threshold $s_0$;
below one can see the non-physical sheet $II$ of the Riemann surface, to which the contour $C^{\prime}$ belonged before the deformation.
The unitary cut delimits the sheets $I$ and $II$. Deformation of the contour $C^{\prime}$ results in residues in the poles of
the logarithmic derivative of $D(s)$, around which the circles are drawn. The arrows indicate direction of circumventing the poles.
}
\label{fig.2}
\end{center}
\end{figure}

A simple heuristic argument illustrating the relationship (\ref {Levinson}) is based on the representation of $D$ function (cf. \cite{Kollins:1977})
\begin{equation} \label{collins}
D(s)=
\prod\limits_{l=1}^{n_{\mathrm{b}}}\frac{s-s_{l}}{s_{0}-s_{l}}%
\prod\limits_{i=1}^{n_{\mathrm{p}}}\frac{s-s_{i}}{s_{0}-s_{i}}
\prod\limits_{j=1}^{n_{\mathrm{CDD}}}\frac{s_{0}-s_{j}}{s-s_{j}}%
\exp \left( -\frac{s-s_{0}}{\pi }\int_{s_{0}}^{+\infty }\frac{\delta (s^{\prime
})-\delta (s_{0})}{(s^{\prime }-s_{0})(s^{\prime } - s + i0)} ds^{\prime }%
\right) .
\end{equation}
In the limit $ s \to - \infty $, the $ D $ function behaves as $ D(s) \sim s^{n_{\mathrm{b}} + n_{\mathrm{p}} - n_{\mathrm{CDD}} + (\delta (+\infty) - \delta (s_{0}))/\pi}$. The standard normalization $D(s) = 1 $ for $s \to -\infty$ is equivalent to (\ref{Levinson}).
%Note that the sign-definiteness condition of the function FP (x) = \ im D (C)$ requires zero modulo $ \ PI$ of the phase difference $ \ Delta (x)-\Delta (s_{0})$ for values $C=s_i$, where primitives are localized.
The finiteness of $N(s) = \Im D(s)$ requires $\delta(s) - \delta(s_{0}) = 0 \mod(\pi)$
for $s=s_j > s_0$ where CDD poles are located.

In the family of $D$ functions, satisfying the Low equation \cite{Low:1955,Krivoruchenko:2010},
$ \Im ( 1/D (s) )$ entering the dispersion integral is bounded.
Zeros corresponding to primitives in the representation (\ref{collins})
match thereby zeros modulo $ \pi $ of the phase difference $ \delta(s) - \delta (s_0)$,
ensuring the sign-definiteness of $N (s)$ in a neighborhood of $s = s_i$.
The constancy of the sign of $N (s)$ is the characteristic feature of belonging $D(s)$ to the class of
generalized $R$ functions which are analytic functions in the complex $s$ plane with the cut $(s_0,+\infty)$,
free of simple zeros outside the real axis \cite{Castillejo:1956}.

Summarizing, the existence of primitives, as well as CDD poles, modifies the Levinson theorem.
Primitives give a negative contribution to the asymptotic value of the scattering phase shift.
In potential scattering, the decrease in the scattering phase with an increase in the energy is associated with a repulsive potential.
In the Dyson model \cite{Dyson:1957} and its generalizations \cite{Krivoruchenko:2010,Krivoruchenko:2011,Krivoruchenko:2011b,Krivoruchenko:2012ii,Krivoruchenko:2017}
the number of CDD poles is determined by the number of bound states, primitives, and resonances ($n_{\mathrm{r}} $):
$n_{\mathrm{CDD}} = n_{\mathrm{b}} + n_{\mathrm{p}} + n_{\mathrm{r}} + \Delta$, where $ \Delta = 0, \pm 1$.
This restriction occurs because between two adjacent CDD poles
there is either a bound state, a resonance, or a primitive.
To enhance repulsion, we could add an additional primitive to the system, but we would have to add another CDD pole to the system.
Primitives and CDD poles enter Eq.~(\ref {Levinson}) with opposite signs, so the asymptotic value of the scattering phase remains unchanged.
Since the value of $ \delta(+\infty) - \delta (s_0)$ is bounded from below, there is a limit to the strength of the repulsive interaction in the model.
Adding new resonances leads to an increase in the $ n_{\mathrm{CDD}} $ and in the asymptotic value of $ \delta(+\infty) - \delta (s_0)$, accordingly.
Such a behavior can be interpreted in terms of an enhanced attraction.
The values of $ n_{\mathrm{r}} $ are not bounded, so unlike repulsion, the attraction in the system can be arbitrarily strong.
The existence of a lower bound of $ \delta (+\infty) - \delta (s_0) $ means
that there exists a stiffest EoS inherent in the $s$-channel exchange models;
this conclusion may be of interest for modeling structure of neutron stars.

In the representation (\ref{collins}), the number of bound states, the number of primitives,
the number of CDD poles and their relative positions on the real axis %of the complex $s$ plane
are arbitrary.
Primitives remove restrictions on the minimum value of $ \delta (+\infty) - \delta (s_0) $,
which can be interpreted to mean the permissibility of arbitrarily strong repulsion between the particles.
The stiffest EoS of nuclear matter obeys thereby requirements of relativistic invariance only,
which include the sound speed limit $ a_s \leq c$. Shock waves carry information and
propagate faster than sound, so there is a more stringent constraint, according to which the speed of propagation
of shock waves in nuclear matter is less than the speed of light.

The family of $D$ functions of Eq.~(\ref{collins}) is greater than the family of $D$ functions of the $s$-channel exchange models.
The same statement could be true for the $t$-channel exchange models.

\vspace{1mm}

The authors are grateful to Yu.~A. Simonov for useful discussions. The present work was performed under the partial support of RFBR Grant 18-02-00733A.

\end{document}